\documentclass[twocolumn]{article}
\pdfoutput=1

\usepackage{array}
\usepackage{dcolumn}
\usepackage{latexsym}
\usepackage{graphicx}
\usepackage{amsfonts}
\usepackage{amsmath}
\usepackage{natbib}
\usepackage{hyperref}

\newcommand{\R}{\mathbb R}

\newcommand{\norm}[1]{\left\Vert#1\right\Vert}

\newcommand{\eps}{\varepsilon}

\newcommand{\DW}{\mathrm{DW}}
\newcommand{\HK}{\mathrm{HK}}

\author{Jan Lorenz\thanks{Universit\"at Bremen, Bibliothekstra\ss{}e, 28359
    Bremen, Germany, math@janlo.de; ETH Zurich, Chair of Systems Design,
  Kreuzplatz 5, 8032 Zurich, Switzerland, jalorenz@ethz.ch}}
\title{Heterogeneous bounds of confidence: Meet, Discuss and Find Consensus!}
\date{First submitted January 3, 2008; revised version August 13, 2009}

\begin{document}
\maketitle

\begin{abstract}
  Models of continuous opinion dynamics under bounded confidence show a sharp transition between a consensus and a polarization phase at a critical global bound of confidence. In this paper, heterogeneous bounds of confidence are studied. The surprising result is that a society of agents with two different bounds of confidence (open-minded and closed-minded agents) can find consensus even when both bounds of confidence are significantly below the critical bound of confidence of a homogeneous society.

  The phenomenon is shown by examples of agent-based simulation and by numerical computation of the time evolution of the agents density. The result holds for the bounded confidence model of Deffuant, Weisbuch  and others (Weisbuch, G. \emph{et al}; Meet, discuss, and segregate!, \textit{Complexity}, 2002, 7, 55--63), as well as for the model of Hegselmann and Krause (Hegselmann, R., Krause, U.; Opinion Dynamics and Bounded Confidence, Models, Analysis and Simulation, \textit{Journal of Artificial Societies and Social Simulation}, 2002, 5, 2).

  Thus, given an average level of confidence, diversity of bounds of confidence enhances the chances for consensus. The drawback of this enhancement is that opinion dynamics becomes suspect to severe drifts of clusters, where open-minded agents can pull closed-minded agents towards another cluster of closed-minded agents. A final consensus might thus not lie in the center of the opinion interval as it happens for uniform initial opinion distributions under homogeneous bounds of confidence. It can be located at extremal locations. This is demonstrated by example.

  This also show that the extension to heterogeneous bounds of confidence enriches the complexity of the dynamics tremendously. 
\end{abstract}

\section{Introduction}
\label{sec:Introduction}

Reaching consensus about certain issues is often desired in a society. In
which society are the chances for consensus better? A society with
homogeneous agents which are equally skeptical about the opinions of
others or a heterogeneous society with open- and closed-minded people? We
study this question in the framework of continuous opinion dynamics under
bounded confidence. The surprising result is that very often a
heterogeneous society can reach consensus even when both open-minded and
closed-minded agents are more skeptical than in a homogeneous society. 

Models of continuous opinion dynamics under bounded confidence have been
introduced independently by Hegselmann and Krause (HK)
\cite{Krause.Stockler1997ModellierungundSimulation,
Krause2000DiscreteNonlinearand,Hegselmann.Krause2002OpinionDynamicsand} 
and Deffuant, Weisbuch and others (DW)
\cite{Deffuant.Neau.ea2000MixingBeliefsamong,
Weisbuch.Deffuant.ea2002Meetdiscussand}. In the basic version of
both models agents adjust their continuous-value opinions toward the
opinions of other agents, but they only take opinions of others into
account if they are closer than a bound of confidence $\eps$ to their own
opinion. The DW and the HK model differ in their communication regime. In
the DW model agents meet in random pairwise encounters. In the HK model
update is synchronous and each agent takes into account only other agents within bounds of
confidence around her current opinion. Both models extend
naturally to heterogeneous bounds of confidence. But the case has only been
briefly addressed in \cite{Weisbuch.Deffuant.ea2002Meetdiscussand} and in 
\cite{Groeber.Schweitzer.ea2009HowGroupsCan} under further extensions to the network of past interactions. 

Opinion dynamics starts with an ensemble of $n$ agents with initial
opinions in the interval $[0,1]$. Dynamics always lead to a stable
configuration where opinions form opinion clusters (see
\cite{Lorenz2005stabilizationtheoremdynamics} for a proof).
If there is only one final cluster the agents have reached consensus. But
there are also other characteristic cluster configurations with respect
to \emph{number}, \emph{sizes} and \emph{locations} of opinion clusters. The
configuration reached is mostly determined by the bound of confidence $\eps$. 
Of course, the final configuration depends also on the specific initial opinions
and in the DW model on the specific realization for the pairwise encounters. But
these parameters are regarded as random and equally distributed in this paper
and in most of the existing literature (with the exception of
\cite{Lorenz.Urbig2007AboutPowerto}). Both models are invariant to joint shifts
and scales of the initial configuration and the bound of confidence. Thus, the
restriction to the opinion interval $[0,1]$ gives still a full characterization of
model dynamics on bounded intervals. 

The simulations of the models show a sharp transition between
a consensus phase and a polarization phase at a critical bound of
confidence. Polarization is meant to be a final state with two equally
sized big clusters while consensus is meant to be a final state where one
big cluster is dominant, usually located in the center of the opinion space.

The behavioral rules of agent-based models can be taken over to density-based
models where dynamics are defined on the density of agents in the opinion
space. This reformulation allows a better estimate of the critical values
for the bound of confidence
\cite{Ben-Naim.Krapivsky.ea2003BifurcationandPatterns,
Fortunato.Latora.ea2005VectorOpinionDynamics,
Lorenz2007ContinuousOpinionDynamics}. The approach also extends
naturally to heterogeneous bounds of confidence.

In Section \ref{sec:models-with-heter} the formal definition of the
agent-based DW and HK model with heterogeneous bounds of confidence
is given and the phenomenon of consensus for low bounds of confidence is demonstrated by
examples. In this paper we only treat two different bounds of confidence,
because this already improves the complexity of the dynamical behavior
significantly. Further on, the density based approach is introduced and
demonstrated by examples. In Section \ref{sec:results} the density-based
approach is used to systematically study the evolving cluster patterns under
uniform initial conditions. Section \ref{sec:drifting} demonstrates the
phenomenon of drifting towards extremes by example. Drifting appear even for slight
unstructured perturbations of the uniform distribution. It is shown by agent-based as well as
density-based examples. Section \ref{sec:conclusions} gives conclusions.

\section{DW and HK model extended to heterogeneous bounds of confidence}
\label{sec:models-with-heter}
In the following we give the definition of both bounded confidence models
including their natural extension to heterogeneous bounds of confidence.

Let us consider $n$ \emph{agents} which hold real numbers between zero
and one as opinions. The \emph{opinion} of agent $i$ at time $t$ is
represented by $x_i(t)$ and $x(t)$ is the vector of opinions of all
agents at time $t$ called the \emph{opinion profile}. Suppose further on,
that agent $i$ has a \emph{bound of confidence} $\eps_i$ which determines
that she takes all agents as serious which differ not more than $\eps_i$
from her opinion. Given an opinion profile $x(t)$ agent $i$ has the
following \emph{confidence set} $I_{\eps_i}(i,x(t)) = \{j \,| \,
\norm{x_i(t) - x_j(t)} \leq \eps_i\}$. The confidence set of agent $i$
contains all agents whose opinions lie in the $2\eps$-interval around
$x_i(t)$. Naturally this includes the agent herself. Agent $i$ with
opinion $x_i(t)$ is willing to adjust her opinion towards the opinions of
others in her confidence set by building an arithmetic mean.

These definitions hold for both models. The differences
of the DW and the HK model comes with the definition of who communicated with whom at what time.

In the DW model two agents $i,j$ are chosen randomly. Each agent changes
her opinion to the average of both opinions if the other agent is in her
confidence set. So
\[
x_i(t+1) = \begin{cases}
\frac{x_i(t) + x_j(t)}{2} & \textrm{if $j \in
  I_{\eps_i}(i,t)$} \\
            x_i(t) & \textrm{otherwise.}
           \end{cases}
\]
The same for $x_j(t+1)$ with $i$ and $j$ interchanged. 

It is important to notice, that it is well possible that one
agent with high bound of confidence changes her opinion, while the
other with low bound of confidence does not. The
convergence parameter $\mu$ of the original model
\cite{Deffuant.Neau.ea2000MixingBeliefsamong,
Weisbuch.Deffuant.ea2002Meetdiscussand} is neglected to not further increase the complexity. 
In \cite{Deffuant.Neau.ea2000MixingBeliefsamong,
Weisbuch.Deffuant.ea2002Meetdiscussand} it has been argued that $\mu$ 
effects only convergence time, later it has been shown that the
parameter also impacts the sizes of minority clusters
\cite{Laguna.Abramson.ea2004Minoritiesinmodel} or might get very important under other extensions \cite{Lorenz.Urbig2007AboutPowerto}. 

\begin{figure}
  \centering
  \includegraphics[width=0.49\columnwidth]{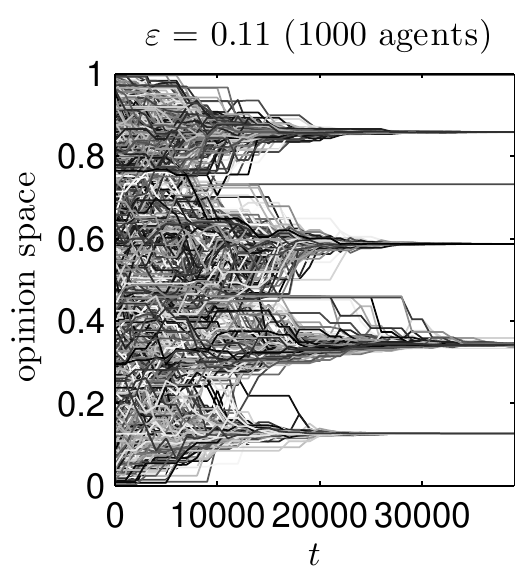}
  \includegraphics[width=0.49\columnwidth]{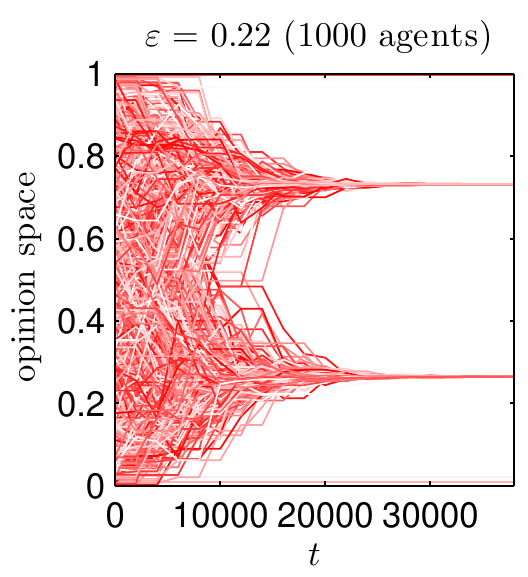}
  \includegraphics[width=\columnwidth]{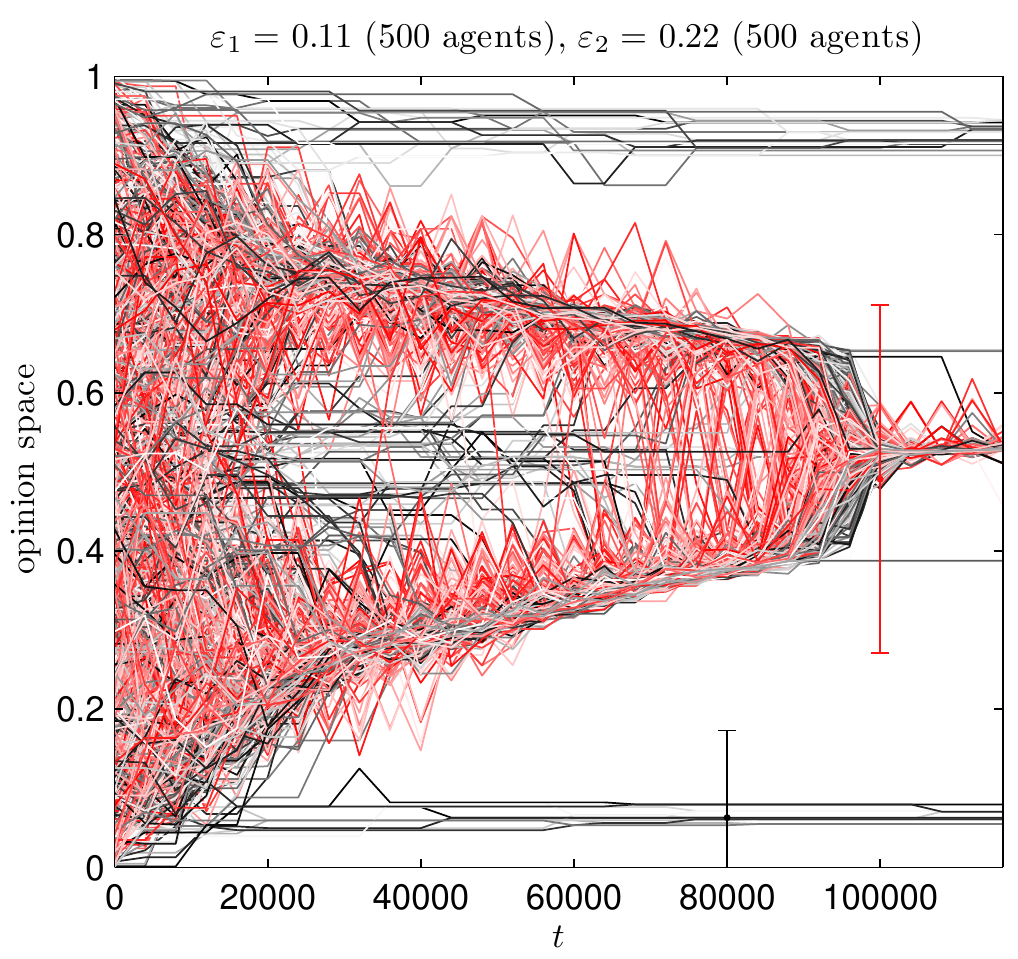} \\
  \caption{DW processes with 1000 agents. Closed minded agents are black, open-minded are (red). Initial conditions in all
    runs are equal. It is possible for the agents to reach consensus under
heterogeneous bounds of confidence (bottom) but not in the corresponding
homogeneous cases (top).}
  \label{fig:abmdw}
\end{figure}

Figure \ref{fig:abmdw} shows an example for the DW model in which 500 of
1000 agents are closed-minded ($\eps_1=0.11$), the other half
open-minded ($\eps_2=0.22$). Both bounds are far less than
$\eps_{\text{crit}}\approx 0.27$ which is the critical bound for the consensus
transition in the homogeneous DW model (see
\cite{Lorenz2007ContinuousOpinionDynamics}). The figure shows the
characteristic patterns with four respectively two big final clusters evolving in 
the cases with homogeneous $\eps_1$ respectively $\eps_2$. The main plot shows how
consensus (neglecting small extremal clusters) is achieved when bounds of confidence are mixed. This is the central
phenomenon this paper is about.

\begin{figure}
  \centering
  \includegraphics[width=0.49\columnwidth]{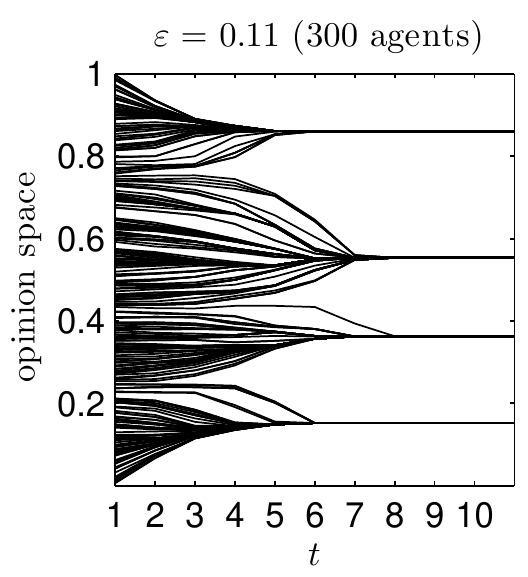}
  \includegraphics[width=0.49\columnwidth]{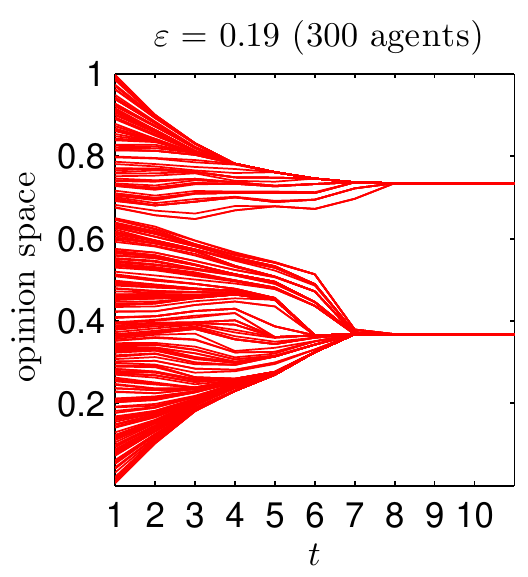}
  \includegraphics[width=\columnwidth]{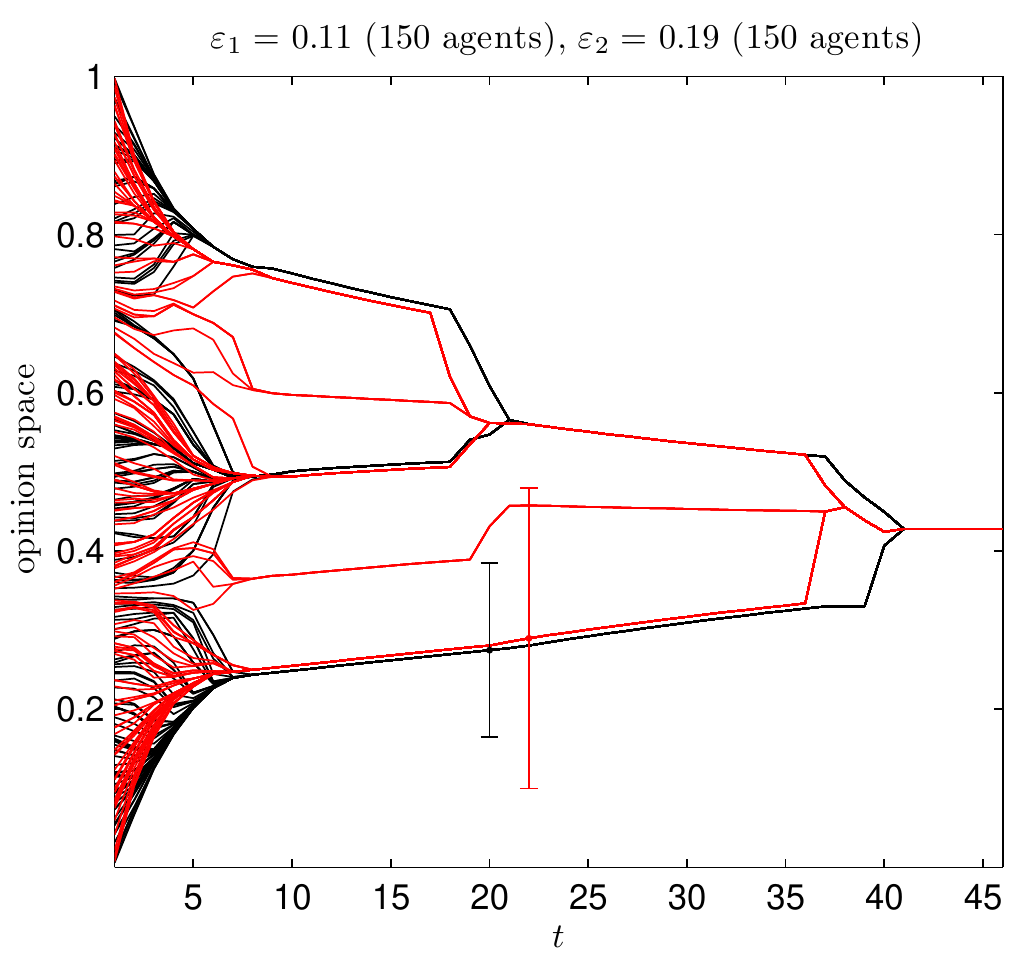} \\
  \caption{HK processes with 300 agents. Closed minded agents are black, open-minded are (red). Initial conditions in all
    runs are equal. Initial conditions in all
    runs are equal. It is possible for the agents to reach consensus under
heterogeneous bounds of confidence (bottom) but not in the corresponding
homogeneous cases (top).}
  \label{fig:abmhk}
\end{figure}

In the HK model all agents act at the same time, and
each changes her opinion to the average of the opinions of all agents in
her confidence set.  So,
\[
x_i(t+1) = \frac{1}{\#I_{\eps_i}(i,x(t))}\sum_{j\in
  I_{\eps_i}(i,x(t))}x_j(t)
\]
for all $i$. ($\#$ is the number of elements of a set.) Figure \ref{fig:abmhk} shows an example for the HK model.
Because of the higher computational effort a small system of 150 closed-minded
($\eps_1=0.11$) and 150 open-minded ($\eps_2=0.19$) is chosen. For the HK model
the critical bound of confidence for the consensus transition lies at about
$\eps=0.19$ (see \cite{Lorenz2007ContinuousOpinionDynamics}), but this result
is achieved using the density based approach and it turns out by simulation
that consensus in the region of 0.19 is only achieved for very
large and very uniform distributed intital conditions. Thus, Figure \ref{fig:abmhk} is another 
example where consensus is never achieved in the homogeneous models (with
system sizes of $n=300$), while mixing can lead to consensus. 

\bigskip

A great success in understanding dynamics of these models was possible
through the introduction of density-based reformulations
\cite{Ben-Naim.Krapivsky.ea2003BifurcationandPatterns} of the DW model. The
basic idea is to define dynamics
on the space of density functions with the opinion interval as the domain
and use the same heuristics as in the agent-based models. So, the scope
changes from a finite number of agents to an idealized infinite number
of agents which are distributed in the opinion interval as defined by the
density function. (In \cite{Blondel.Hendrickx.ea20072Rconjecturemulti-agent}
another model with an infinite number of agents is introduced which allows to
transform agent-based dynamics more straight forward.  The density-based
approach can be derived from that.) This way, the evolution of the agent
density in the opinion interval can be computed numerically for any initial
opinion density. This allows to get an overview about the average
behavior of agent-based dynamics by computing numerically just one
evolution of the agent density in the opinion interval. The density-based
computation matches agent-based fairly well when
the number of agents is sufficiently large and the initial
agent-based opinion profile is a proper draw from the initial agent distribution in
the density-based model. Thus, this approach avoids noisy Monte-Carlo
simulations and gives a good overview on attractive states.
The density-based approach has been used to derive bifurcation diagrams
for the evolving cluster patterns with respect to a homogeneous bound of
confidence for the initial opinion distributions to be uniform in the opinion
space. 

For the HK model the same approach was first applied independently in
\cite{Fortunato.Latora.ea2005VectorOpinionDynamics} and
\cite{Lorenz2005Continuousopiniondynamics}, for an overview and discussion of
different methods see \cite{Lorenz2007ContinuousOpinionDynamics}. 


Density-based models have been proposed in continuous time and continuous
opinion space
\cite{Ben-Naim.Krapivsky.ea2003BifurcationandPatterns,
Fortunato.Latora.ea2005VectorOpinionDynamics} as well as in
discrete time and opinion space
\cite{Lorenz2005Continuousopiniondynamics,
Lorenz2006Consensusstrikesback,Lorenz2007RepeatedAveragingand}. The
discrete version takes the form of an interactive Markov chain, where
transition probabilities depend on the current state of the system. For
numerical computation the continuous opinion interval as well as time has
to be discretized anyway. Therefore, we take the discrete approach
directly in this paper.

The density-based models with homogeneous bounds of confidence can be
extended for heterogeneous bounds of confidence straight forward by
introducing a density function for each bound of
confidence. The precise definition follows for simplicity for just two
bounds of confidence $\eps_1$ and $\eps_2$. This setting is what we
simulate systematically in the next section. The definition can be easily
extended to more bounds of confidence and even a continuum of bounds of
confidence.

Instead of agents and their opinions we define the state of the
system as a density function on the opinion interval which evolves in
time.  We discretize the opinion space $[0,1]$ into $n$ subintervals
$[0,\frac{1}{n}[, [\frac{1}{n},\frac{2}{n}[,\dots,[\frac{n-1}{n},1]$
which serve as \emph{opinion classes}. So, we switch from $n$ agents with
opinions in the opinion interval to an idealized infinite population,
which is divided into $n$ opinion classes. We label opinion classes with $\{1,\dots,n\}$ such that class $1\leq i\leq n$ stands for opinions in the
interval $[\frac{i-1}{n},\frac{i}{n}[$. The two bounds of confidence $\eps_1$ and
$\eps_2$ naturally transform with respect to the $n$ opinion classes
to their discrete counter-parts $\epsilon_1
= n\eps_1$ and $\epsilon_2 = n\eps_2$.

The state of the system is quantified by two row vectors $p^1(t),p^2(t) \in
\R^n_{\geq 0}$, where $p_i^k(t)$ is the fraction of the total
population which holds opinions in class $i$, have a bound of confidence
$\epsilon_k$ at time $t$. The pair $(p^1(t),p^2(t))$ is called the \emph{opinion
distribution} at time $t$. One can see each vector $p^k$ as the histogram of
the agents with bound of confidence $\epsilon_k$ over the opinion classes.
Naturally, it should hold that the fractions sum up
to one $\sum_{i,k} p^k_i(t) = 1$. Further on, we define $p(t) =
p^1(t) + p^2(t)$ to be the opinion distribution of the full population
regardless of the bounds of confidence. We choose row vectors $p^k$ because this
is a convention in defining discrete Markov chains. A discrete Markov chain is
given by its transition matrix $T$ where $T_{ij}$ is the probability that an
agent switches from state $i$ to $j$. Given a distribution $p(t)$ the
next time step's distribution is thus computed by $p(t+1) = p(t) T$. In our
case the transition matrices will be a function of the current state. This
is called an interactive Markov chain. 

We consider that agents never change their bound of confidence. Therefore,
we can define a transition matrix for the agents with discrete bound of confidence
$\epsilon_k$ as $T(p(t),\epsilon_k)$. Notice that the transition matrix depends on the
opinion vector for the full population $p(t) = p^1(t) + p^2(t)$ only. This reflects,
that the change of opinion of an agent with bound of confidence $\epsilon_k$
does not depend on the bounds of confidence of the other agents, only on the
distribution of opinions in total and its own bound of confidence. The
density-based dynamics is then
defined for the initial distributions $(p^1(0), p^2(0))$ as
\begin{align}
  p^1(t+1) = p^1(t) T(p(t),\epsilon_1) \\
  p^2(t+1) = p^2(t) T(p(t),\epsilon_2).
\end{align}

This framework is applied for both models. So, we have to specify the
transition matrices for the two models now. 

The \emph{Deffuant-Weisbuch transition matrix} is defined by
\[
T^\DW_{ij}(p,\epsilon) = \left\{%
  \begin{array}{ll}
    \frac{\pi^i_{2j-i-1}}{2} + \pi^i_{2j-i} + \frac{\pi^i_{2j-i+1}}{2}, & \hbox{if $i\neq j$, } \\
    q_{i}, & \hbox{if $i=j$.} \\
  \end{array}%
\right.
\]
with $q_i = 1 - \sum_{j\neq i, j=1}^n T^\DW_{ij}(p,\epsilon)_{ij}$ and
\[\pi^i_m = \left\{%
  \begin{array}{ll}
    p_m, & \hbox{if $|i-m|\leq \epsilon$} \\
    0, & \hbox{otherwise} \\
  \end{array}%
\right.\]
For $i<1$ and $i>n$ it is defined $p^k_i = 0$ for convenience.
The probability of an agent to change from opinion $i$ to opinion $j$
depends on the fractions of agents in the opinion classes $2j-i-1,2j-i,2j_i+1$,
but only when these classes are not farther than $\epsilon_i$ from $i$.
The average of $i$ and $2j-i$ is indeed $j$. The average of $i$ and $2j-i-1$ i
s $j-\frac{1}{2}$, thus only half of the agents is expected to switch to state
$j$ (the other half will switch to state $j-1$). Analog for averaging $i$ and
$2j-i+1$.

\begin{figure}
  \centering
  \includegraphics[width=\columnwidth]{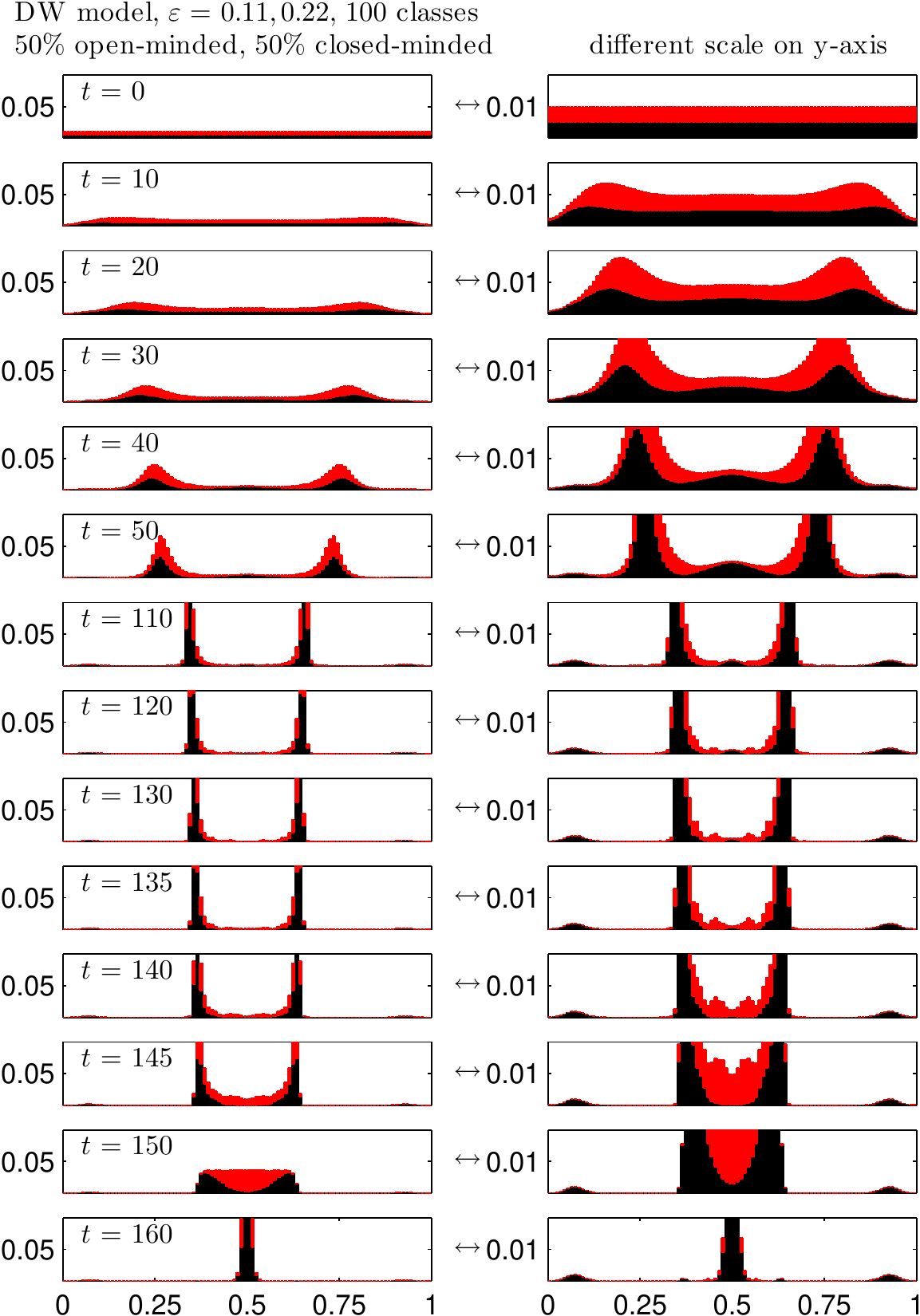}
  \caption{A density-based DW process with closed minded (black) and
open-minded (red)  agents. Time proceeds downwards. Notice that time steps
are not equidistant, but selected to show important changes. The right
hand side is just another scale of the y-axis which makes small classes
visible. Values of bounds of confidence coincide Figure \ref{fig:abmdw}. Also a consensus is
found (neglecting a small proportion of extremists). }
  \label{fig:dwcons}
\end{figure}

Figure \ref{fig:dwcons} shows an example computation for 
the evolution of an opinion distribution in the DW model. The distributions
of the open-minded population
(red) and the closed-minded (black) population are stacked to give an
impression of the evolution of the whole population. Computations where
carried out with 100 opinion classes, but runs would look essentially
identically for a higher number of classes and an appropriate scaling of
the discrete bounds of confidence. The example is with the same parameters as
the agent-based example in Figure \ref{fig:abmdw}, and indeed shows the same
phenomenon of convergence to a big central consensual cluster due to the interplay
of closed and open-minded agents. Both groups play its role in reaching
consensus. The closed-minded ensure that intermediately ($t=50$) there is a
small cluster in the center of the opinion space. Open-minded agents at the
same location would already been absorbed by the two big intermediate clusters
on the right and on the left. Finally, the open-minded play their role in
pulling the closed-minded from both sides slowly towards the center. 

\bigskip

The \emph{HK transition matrix} is defined by 
\[
T^\HK_{ij}(p,\epsilon) = \left\{%
  \begin{array}{ll}
    1 & \hbox{if $j = M_i$,} \\
    \lceil M_i\rceil - M_i & \hbox{if $j = \lfloor M_i \rfloor$, $j\neq M_i$,} \\
    M_i - \lfloor M_i\rfloor & \hbox{if $j = \lceil M_i \rceil$, $j\neq M_i$,}  \\
    0 & \hbox{otherwise.} \\
  \end{array}%
\right.
\]
with 
\[
M_i(p,\epsilon) =
\frac{\sum_{k=i-\epsilon}^{i+\epsilon}kp_k}{\sum_{k=i-\epsilon}^{i+\epsilon}
p_k}
\]
being the \emph{$\epsilon$-local mean} at opinion class $i$. The brackets
$\lfloor\cdot\rfloor$ represents rounding to the upper integer,
$\lfloor\cdot\rfloor$ rounding to the lower integer. The
$\epsilon$-local mean is the barycenter of distribution $p$ on the discrete
interval of length $2\epsilon$ centered on opinion $i$. So, an agent switches
from opinion $i$ to $j$ when $j$ is the $\epsilon$-local mean of $i$. If the
$\epsilon$-local mean is not an integer it switches to the class above or below
with probabilities depending on the distance of the $\epsilon$-local mean to
these classes. 

\begin{figure}
  \centering
  \includegraphics[width=\columnwidth]{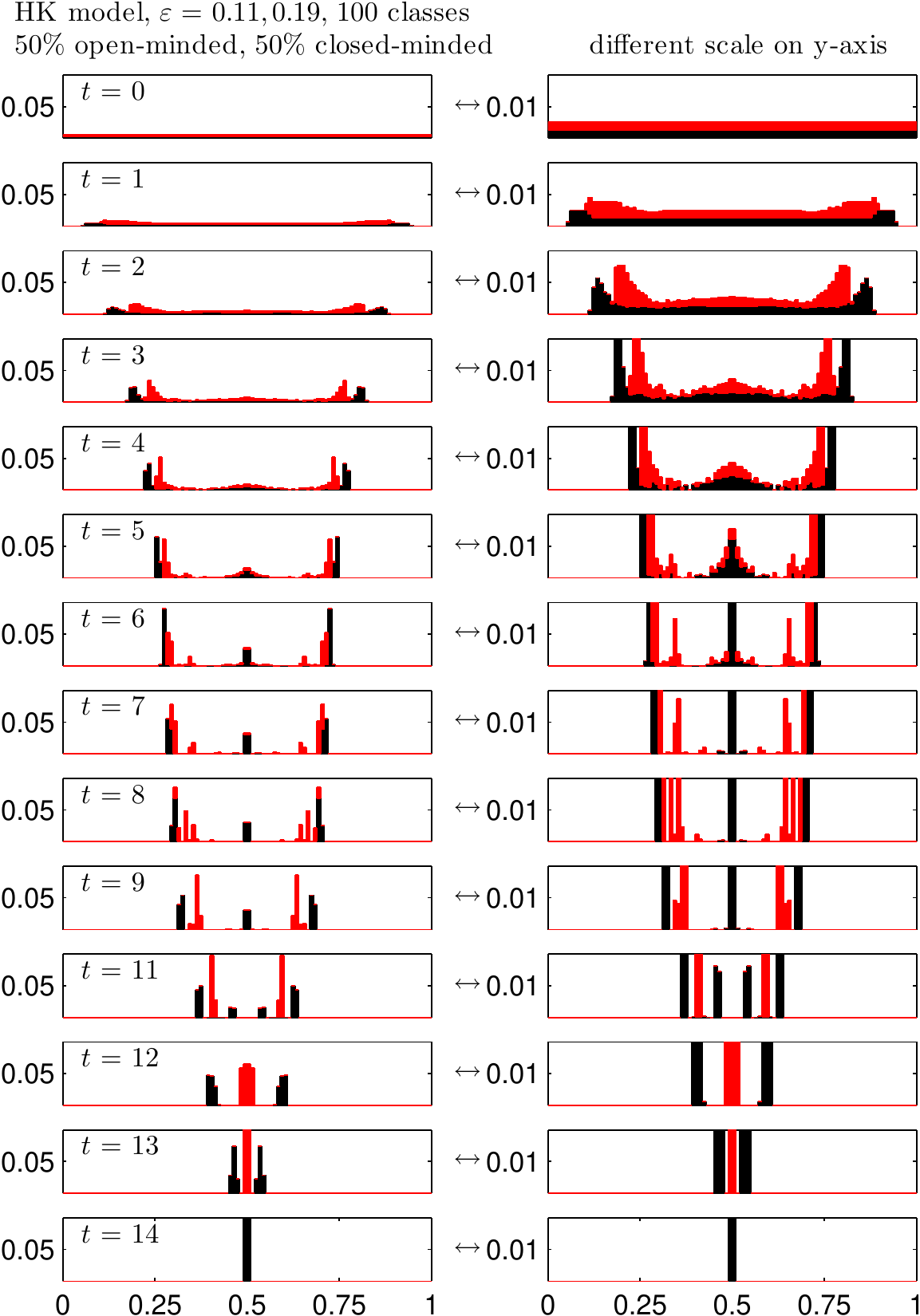}
  \caption{A density-based HK process with closed minded (black) and
open-minded (red)  agents. Time proceeds downwards. Notice that time steps
are not equidistant, but selected to show important changes. The right
hand side is just another scale of the y-axis which makes small classes
visible. Values of bounds of
    confidence coincide Figure \ref{fig:abmhk}. Also a consensus is
found. }
  \label{fig:hkcons}
\end{figure}

Figure \ref{fig:hkcons} shows an example computation for 
the evolution of an opinion distribution in the HK model, which is in the same
style of presentation as Figure \ref{fig:dwcons} and matches the parameters of
Figure \ref{fig:hkcons}. Again, consensus is found only due to the interplay of
the closed- and the open-minded. A remark on comparison to the bifurcation
diagram for the HK model reported in \cite{Lorenz2007ContinuousOpinionDynamics}
which gives $0.19$ as the critical value of the consensus transition. But
consensus in this region is only achieved under very long convergence time and
only for a large number of classes, there 1000, which corresponds to a very
large number of agents in the agent-based version. So, consensus is possible in
very large homogeneous groups under $\varepsilon=0.19$ by very long convergence
times. In the example of Figure \ref{fig:hkcons} consensus is achieved very
fast under the heterogeneous bounds of confidence. 

The discrete time and discrete opinion space approach for density-based
 models presented here and the continuous approaches (DW
\cite{Ben-Naim.Krapivsky.ea2003BifurcationandPatterns}, HK
\cite{Fortunato.Latora.ea2005VectorOpinionDynamics}) for density-based models
have been shown to lead to the same results for the DW model
\cite{Lorenz2007RepeatedAveragingand}. This does
not hold for the HK
model (see \cite{Lorenz2007ContinuousOpinionDynamics} for a discussion). Further
on, it matters for the HK model if the discretization of the opinion interval
is into an even or odd number of bins (see
\cite{Lorenz2005Continuousopiniondynamics} for
evidence). In the following section we focus on odd numbers. Thus on
distributions where a central bin exists which is the natural candidate for a
consensus under a symmetric initial distribution. In the following section we
will show  by systematic simulations that the phenomenon of reaching
consensus with lower but heterogeneous bounds of confidence is generic in
the both models.

\section{Systematic Simulation}
\label{sec:results}
In this section a complete picture is given about the final `degree of consensus'
for societies of closed- and open-minded agents which initial
opinions are uniformly distributed in the opinion interval. Only the case of
equally sized groups of closed- and open-minded agents is treated. 

\begin{figure}[htbp]
  \centering
  \includegraphics[width=1.00\columnwidth]{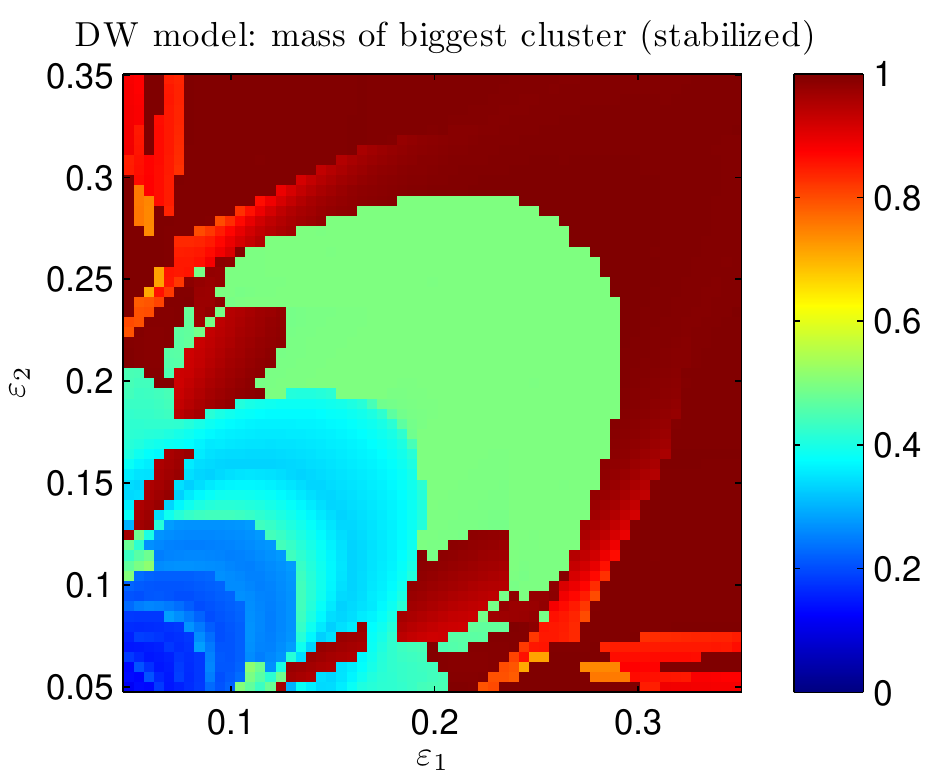}\\
  \includegraphics[height=0.4\columnwidth]{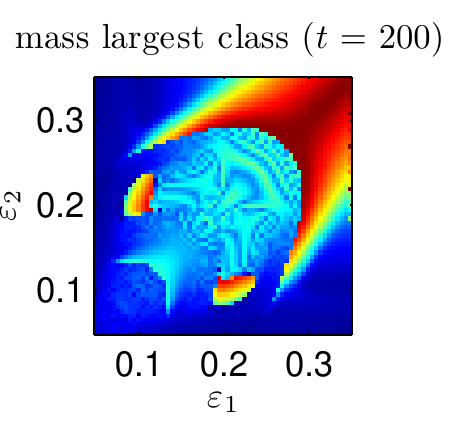}
  \includegraphics[height=0.4\columnwidth]{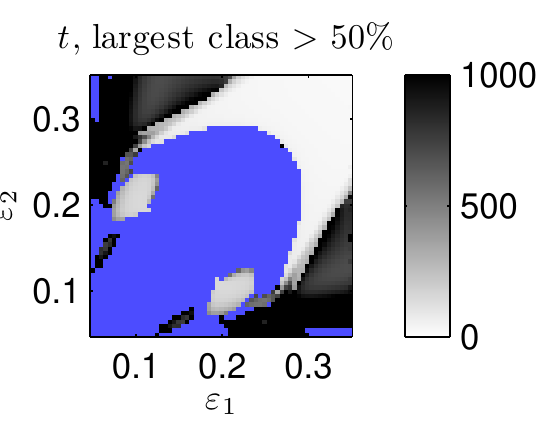}
  \caption{Mass of the biggest cluster with precision $10^{-4}$
after stabilization for the DW model (top). Mass of the largest class at time
step $t=200$ (bottom left), and the time step $t$ when the mass of the central
cluster exceeds 50\% (blue stands for `never').}
  \label{fig:phaseDW}
\end{figure}

For a systematic simulation setup, the opinion space $[0,1]$ is divided into
$n=201$ opinion classes. The initial opinion distribution is
$p^k_i=\tfrac{1}{402}$ for $k=1,2$ and $i=1,\dots,n$. Then the final opinion
distribution is computed for $\epsilon_1,\epsilon_2 = 10,11,\dots,70$ which
corresponds to $\eps_1,\eps_2 = \frac{10}{201},\dots,\frac{70}{201}\approx
0.05,\dots,0.35$. Notice that a formal proof for convergence to a stable opinion
distribution is still lacking, see \cite{Lorenz2007ContinuousOpinionDynamics} for
discussion. For our setting convergence is evident by observation, but stopping
criteria for simulation runs are difficult to define, especially for the DW
model. For the HK model new time steps were computed until the difference to the
former time steps got zero (due to computer precision limits). Convergence was
achieved in reasonable time. Numerical problems evolved when distributions got
asymmetric around the central class 101 for the reason of floating point
errors. These problems were circumvented by making the opinion distributions
symmetric again after each iteration. For the heterogeneous DW model stopping
criteria are more complicated because it has a rich variety of types of
convergence which are not fully classified and understood until now.  Further
on, convergence can last very long and it is difficult to decide whether
convergence will lead to another drastic change once or not.  Therefore, 
three different ways to visualize the results are chosen in Figure \ref{fig:phaseDW}.

To quantify the `degree of consensus' the \emph{mass
of the biggest cluster} is an appropriate measure. In a stabilized final
opinion distribution a cluster is a set of at most two adjacent classes with
positive mass surrounded by classes with zero mass (see
\cite{Lorenz2007RepeatedAveragingand,Lorenz2007Fixedpointsin}). Due to the
odd number of classes and symmetry a cluster including the central class $i=101$
can finally only be a one-class cluster. The central class $i=101$ is also the
only candidate where $p^1_i+p^2_i>0.5$ is possible due to conservation
of symmetry. Convergence in the DW model is slow. This, concerns especially the
final condensing to clusters, even when the broad separation into clusters has settled. Further
on, masses remain always positive (though small). Thus, we have to come up with a
cluster definition which we can apply in not fully converged situations. So we
define an opinion cluster with precision $10^{-4}$ to be a set of adjacent
opinion classes which all contain mass larger than $10^{-4}$ and which
neighbor classes have mass less than $10^{-4}$. 

\begin{figure}[htbp]
  \centering
  \includegraphics[width=1.00\columnwidth]{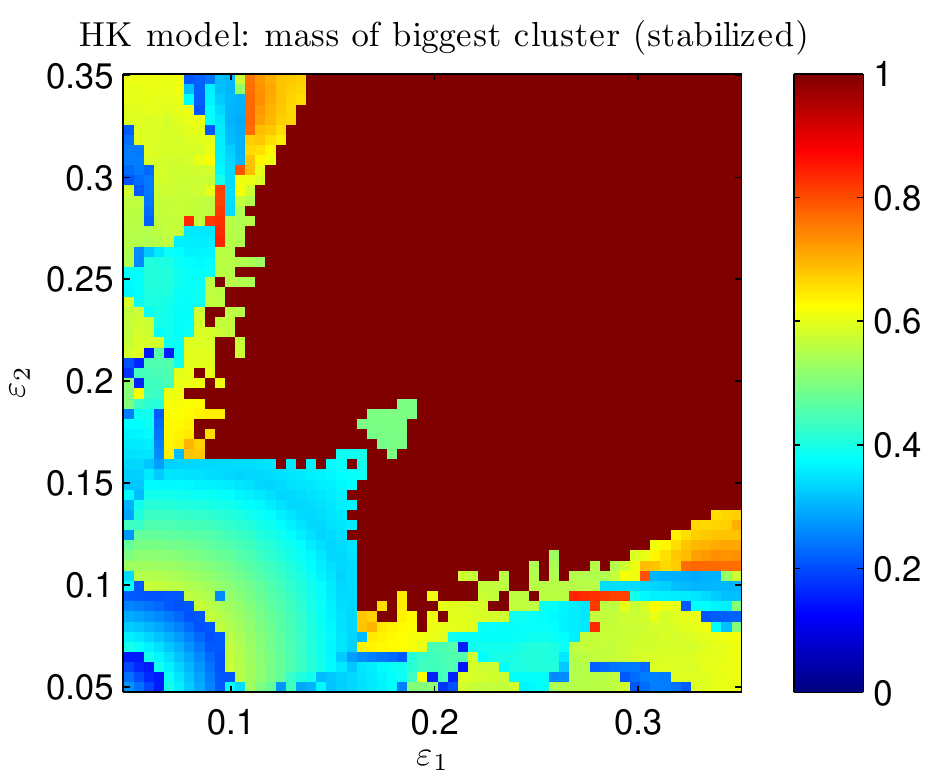}
  \caption{Mass of the biggest cluster after
stabilization for HK model.}
  \label{fig:phaseHK}
\end{figure}

The masses of the biggest cluster are documented in Figure
\ref{fig:phaseDW} for the DW model and in Figure \ref{fig:phaseHK} for
the HK model. We color the plane of all $(\eps_1,\eps_2)$ points with the
values of the mass of the biggest cluster after stabilization. So, regions of
certain degrees of consensus are dark red, while regions of almost consensual 
clusters are orange and red. Several abrupt and continuous transitions in
changes of $\eps_1,\eps_2$ are visible. 

Let us first take a look on the diagonal of the plots which represents
the homogeneous situation $\eps_1=\eps_2$. We see the points of the
consensus transition as about $0.27$ for the DW model and about $0.19$
for the HK model. (It can also be observed that consensus strikes back
for $0.17$ in the HK model. See \cite{Lorenz2006Consensusstrikesback} for
details about this phenomenon.) The surprise comes when we look on the
heterogeneous situations besides the diagonal. Consensus is in many cases
possible even when $\eps_1$ and $\eps_2$ are both below the critical $\eps$ for
the consensus transition in the homogeneous case.

The evolution of consensus under heterogeneous bounds of confidence in
the DW model looks quite ubiquitous. But it is important to notice that
the plot for the DW model might look very different for different levels
of precision. Further on, convergence time to consensus could be very
slow. To clarify the picture about the DW model the two smaller plots are 
included in Figure \ref{fig:phaseDW}. The bottom left plot shows the situation
after a not so long time $t=200$. Because clusters can not be determined at this
level we simply plot the maximum of $p_i(200)$ all
opinion classes $i\in\{1,\dots,n\}$. One can see that especially in the region
around $(\eps_1,\eps_2) = (0.11,0.22)$ the maximum has already exceeded
$50\%$.  So, in this region consensus will appear after a reasonable
amount of time. The bottom right plot shows the time steps when the central
class contains more than $50\%$ of the mass (if this
happens at all). So, this is another measure for convergence to consensus
in a reasonable time. The color axis has been adjusted from zero to
thousand. For the blue area the central cluster has not exceeded
$50\%$ either because of convergence to polarization or plurality (in the
region around the diagonal) or because of too slow convergence (in the region at
the corners). So, black stands for a huge time interval. The nonconverged
states in the upper left and bottom right corner did not converged after
$45.000$ time steps. Further on, there is one interesting example
for long convergence time on the diagonal (the homogeneous case). This is an
example for a very short $\eps$-phase where convergence to consensus in the DW
model is via a metastable polarized state, which has not been observed before.
This slow convergence $\eps$-phase close to the transition is really huge for
the HK model (see \cite{Lorenz2006Consensusstrikesback}).

We can conclude that enhancement of the chances for consensus due to
heterogeneity is generic. It appears due to a subtle interplay between closed-
and open-minded agents. The results presented were all produced for initial
opinion distributions which were perfectly uniform. In the following section we
will study perturbations of this situation.

\section{Inherent drifting towards the extremes}
\label{sec:drifting}

\begin{figure}
  \centering
  \includegraphics[width=\columnwidth]{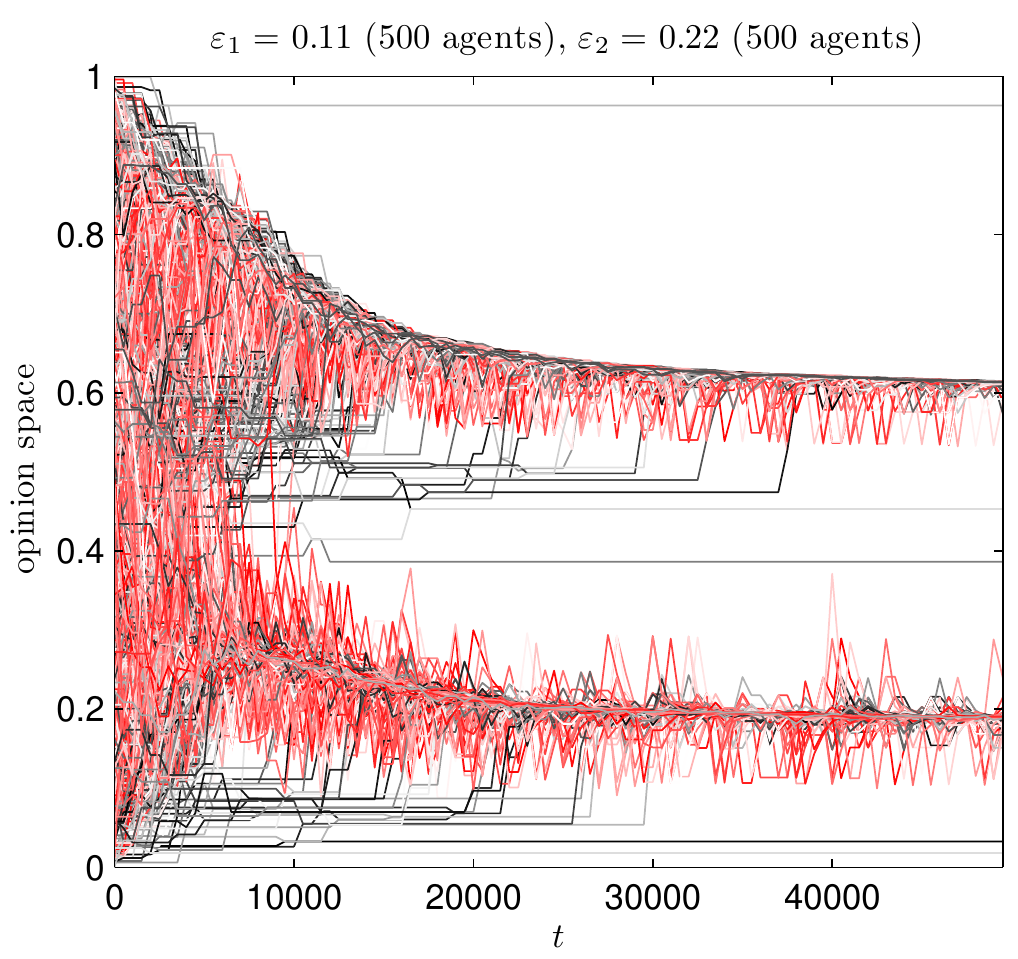}
  \includegraphics[width=\columnwidth]{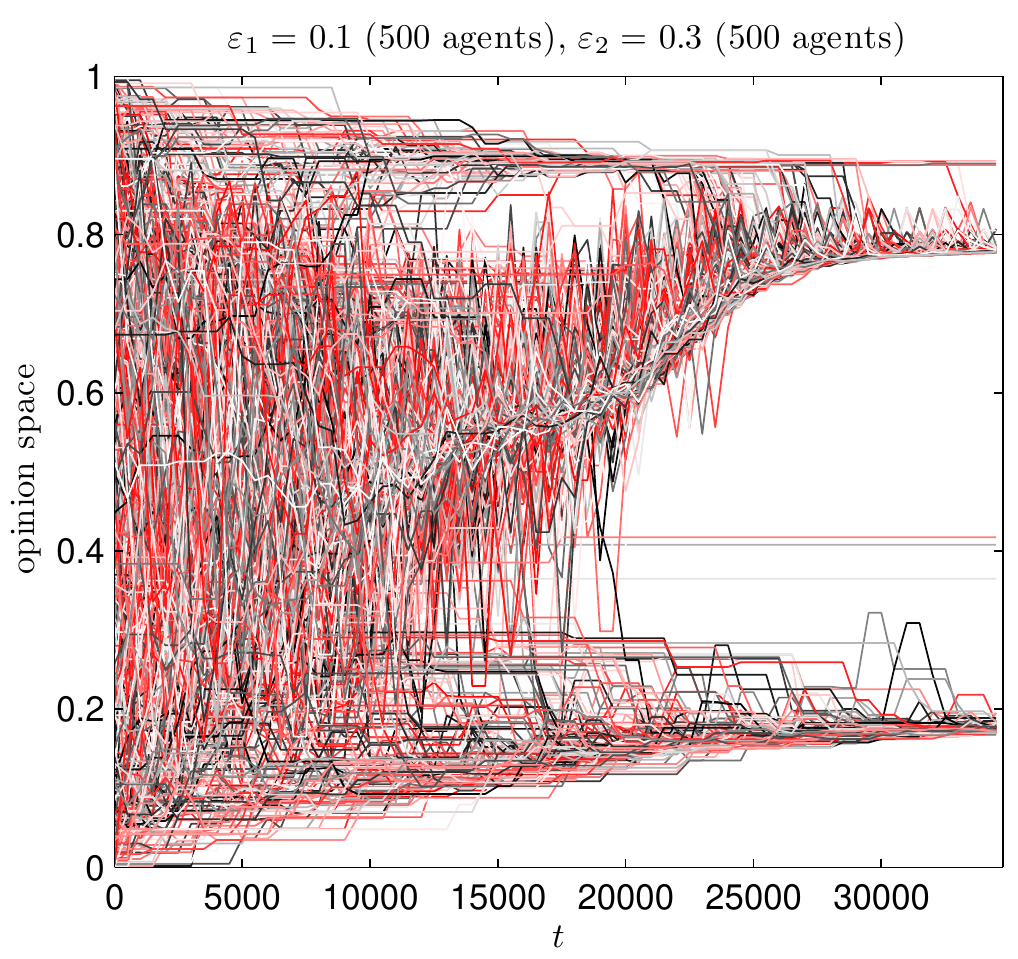}
  \caption{Two DW processes with 1000 agents. The top plot is another realization 
of Figure \ref{fig:abmdw} with a different realization of the random pairwise
encounters. The bottom plot is an example for an extremal consensus under other parameters.}
  \label{fig:hetdwdrift}
\end{figure}

In the following it is demonstarted by example that even unstructured deviations
from the perfect symmetry in the initial opinion distribution can have important
consequences. Let us start by looking on some other agent-based examples.
Figure \ref{fig:hetdwdrift} shows two different runs  The upper plot shows a
simulation run for the same initial conditions as Figure \ref{fig:abmdw} but
with a different realization of the pairwise encounters. Consensus is not
achieved but a polarization into two big clusters. Moreover, both cluster drift
towards zero. This happens in both big clusters due to their contact to a very small group of
closed-minded agents which lies below each of them. This situation was achieved in the first
10000 time steps by the upper group moving slightly faster towards the center and
thus attracting more of the closed-minded agents in the center. This brought the
emerging big lower cluster to be more oriented to the even lower small cluster
of closed minded. This established the situation of the overall drift to one
side. Notice that this situation was not put a priori in, but emerged from a
uniform distribution of open-minded as well as closed-minded agents. 
The bottom plot in Figure \ref{fig:hetdwdrift} shows an example with different
parameters, which leads to an extremal consensus close to one. Again the
initial configuration is unstructured. This shows that the system becomes suspect
to extremism even when this susceptibility is not visible a priori. Extremism
in bounded confidence models has already been studied
\cite{Deffuant.Neau.ea2002HowCanExtremism,Amblard.Deffuant2004roleofnetwork,
Deffuant2006ComparingExtremismPropagation}, but in all of these studies the
susceptibility to extremism was put in a priori by populating the extremes with
very closed-minded agents, such that they form natural attractors for the
open-minded central agents. The question in these studies was then just: under
what conditions do central agents convergence in the center, split towards the two extremes
or convergence together to one extreme. In this study, we see that also the situation
which is susceptible to extremal convergence can emerge dynamically.

\begin{figure}
  \centering
  \includegraphics[width=\columnwidth]{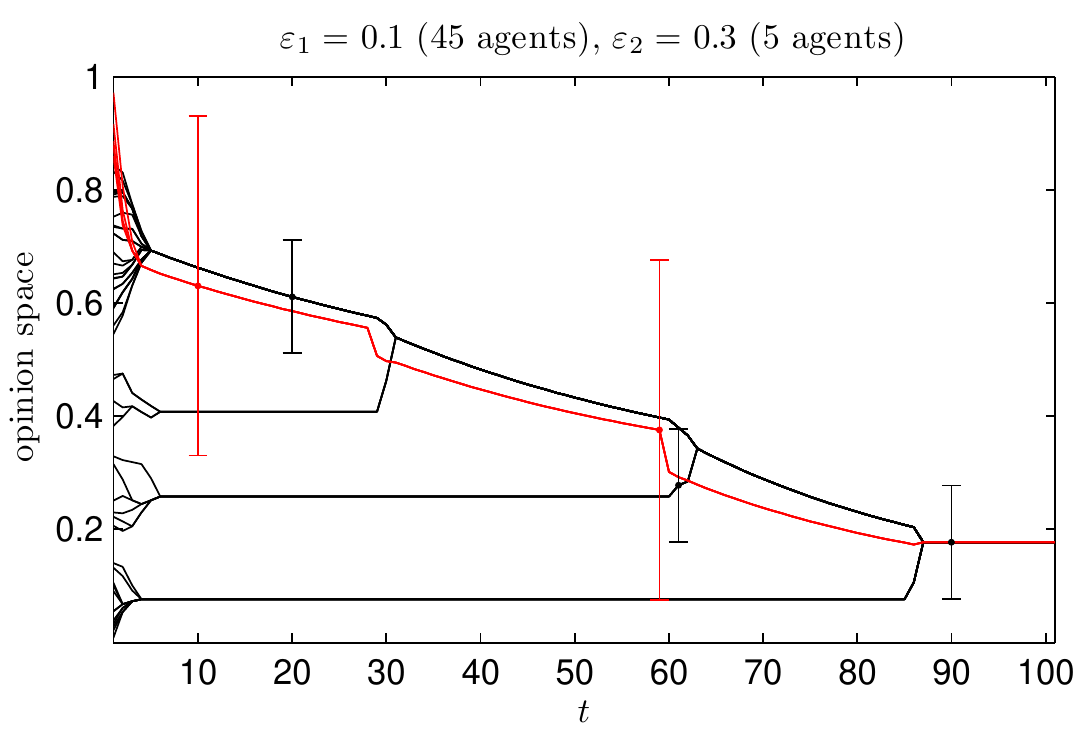}
  \includegraphics[width=\columnwidth]{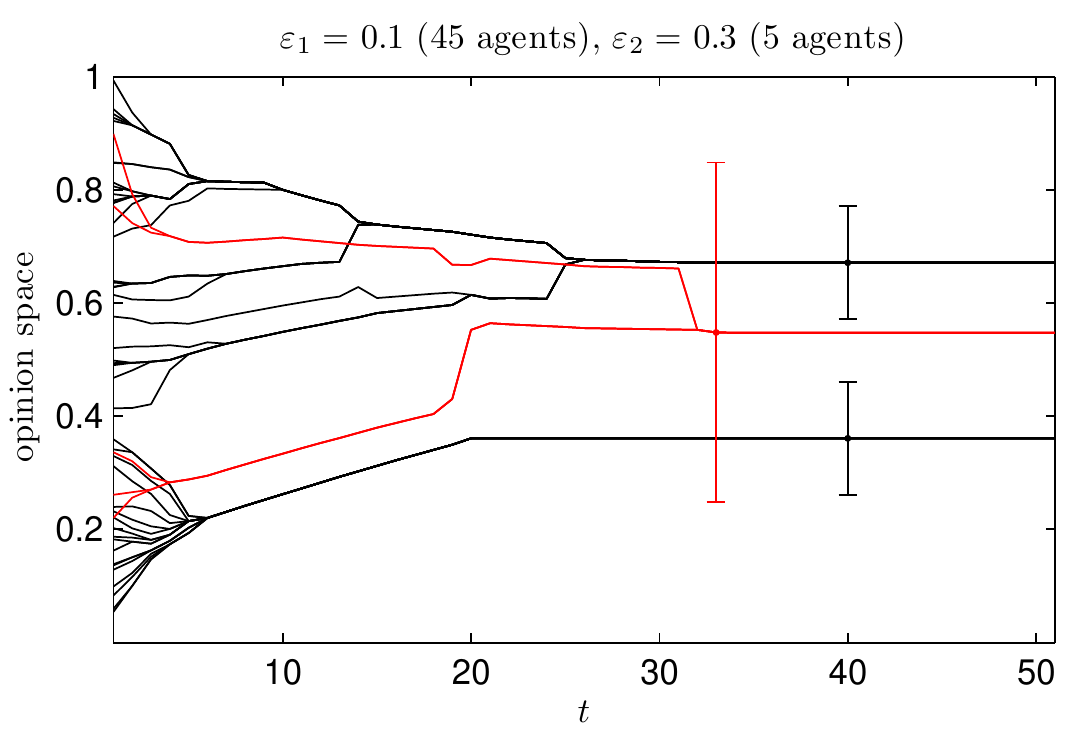}
  \caption{Two HK processes with 50 agents, where only few are
    open-minded (red) and the majority is closed-minded (black). The top plot
is an example of extremal consensus, the bottom plot for an evolving sitting between
the chairs situation.}
  \label{fig:hethkdrift}
\end{figure}

Figure \ref{fig:hethkdrift} show two example runs for the HK model. In these
examples 90\% of the population is closed-minded and 10\% is open-minded. This
is chosen because the 50\%/50\% situation does not show many variations, as
should be shown in this section. The upper plot shows an impressive example of
convergence to consensus caused by only 10\% of open-minded agents but for the
cost that the consensus is a very extreme opinion. Moreover it is the opposite
of the opinion the open-minded agents started with. The lower plot starts with
the same initial profile of opinions but with a different choice of
the five open-minded agents. The system converged to a frozen situation, where the
open-minded agents finally sit between the chair. They take the opinions of both closed-minded
clusters into account, but are not able to pull them together. 

\begin{figure}
  \centering
  \includegraphics[width=\columnwidth]{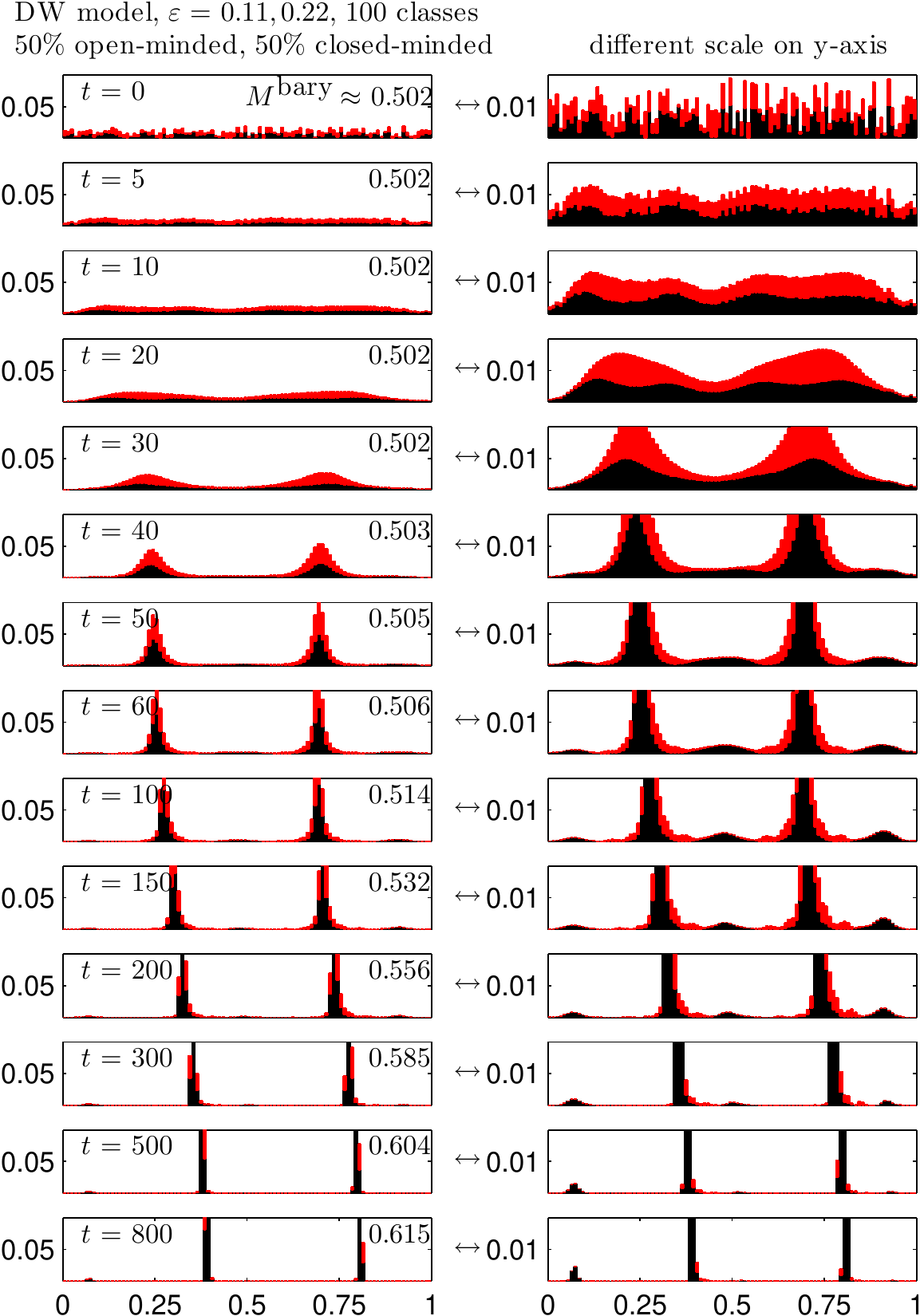}
  \caption{The situation of Figure \ref{fig:hetdwdrift} (top) in the density based
model. Also comparable to Figure \ref{fig:dwcons} with a perturbed initial
distribution. $M^\textrm{bary}$ indicates the overall average opinion, which
quantifies the overall drift.}
  \label{fig:dbmdwdrift}
\end{figure}

\begin{figure}
  \centering
  \includegraphics[width=\columnwidth]{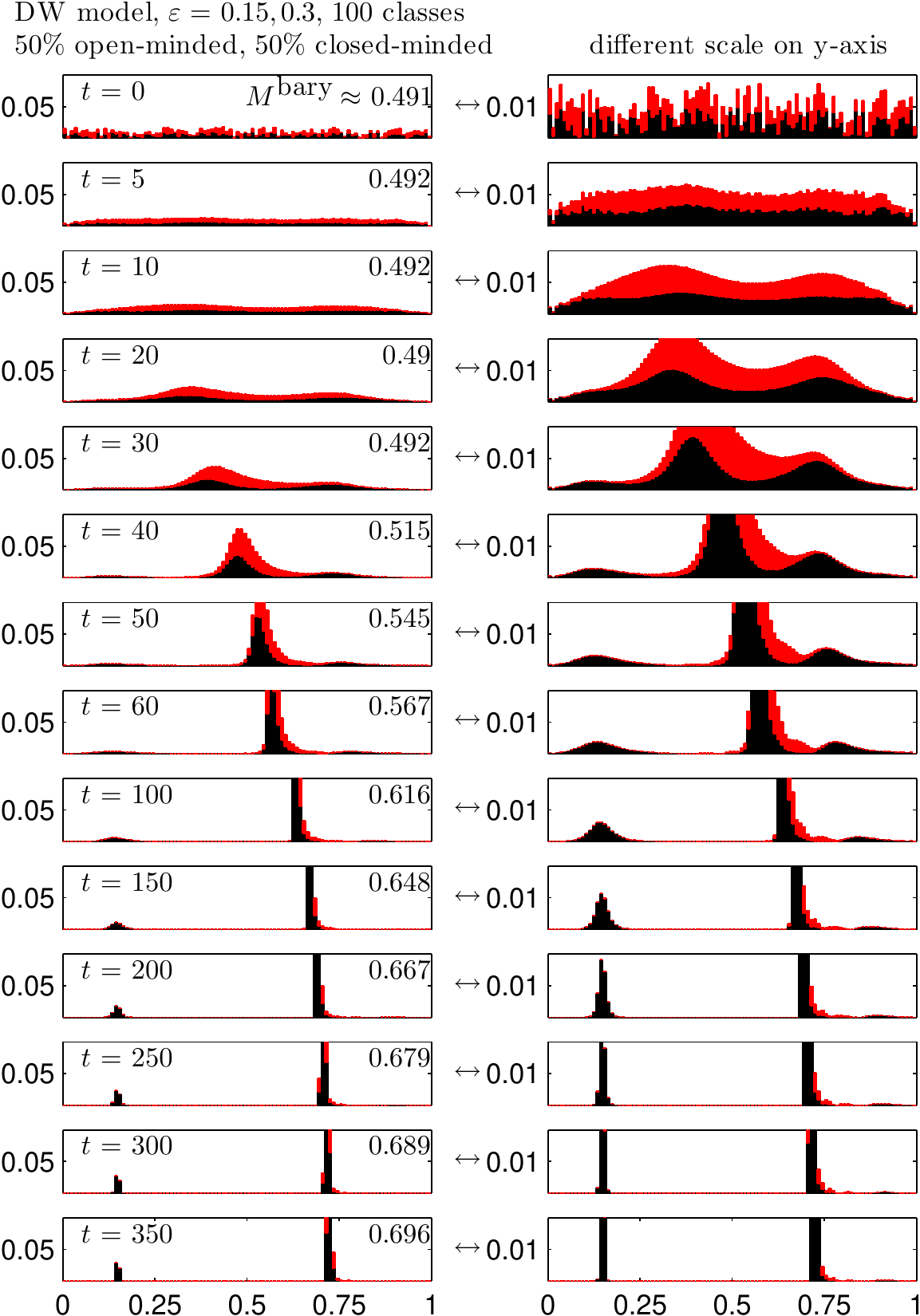}
  \caption{The situation of Figure \ref{fig:hetdwdrift} (bottom) in the density based
model. $M^\textrm{bary}$ indicates the overall average opinion, which
quantifies the overall drift. }
  \label{fig:dbmdwdriftextreme}
\end{figure}

\begin{figure}
  \centering
  \includegraphics[width=\columnwidth]{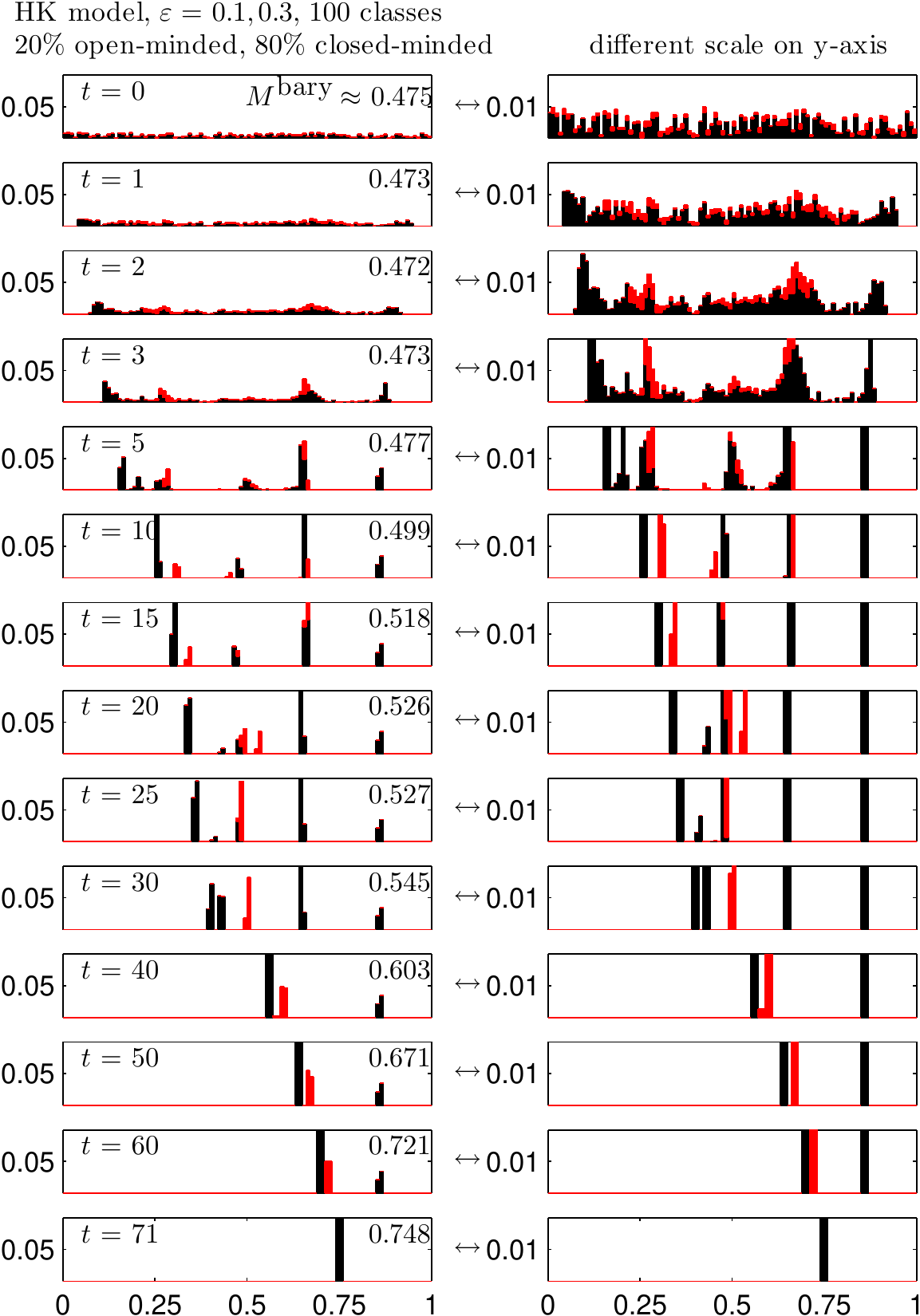}
  \caption{The situation of Figure \ref{fig:hethkdrift} (top) in the density based
model. $M^\textrm{bary}$ indicates the overall average opinion, which
quantifies the overall drift. .}
  \label{fig:dbmhkdrift}
\end{figure}

\begin{figure}
  \centering
  \includegraphics[width=\columnwidth]{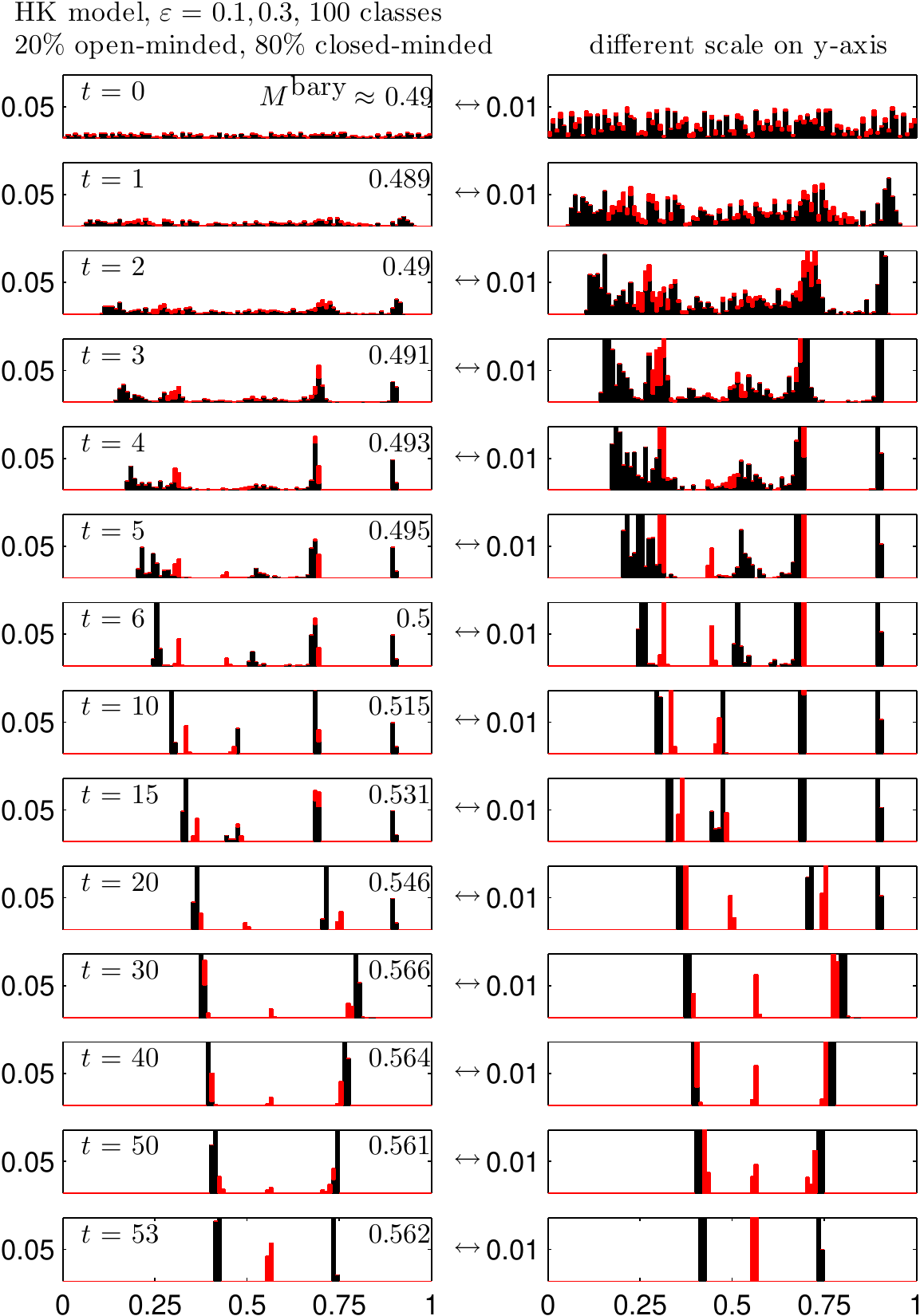}
  \caption{The situation of Figure \ref{fig:hethkdrift} (bottom) in the density based
model. $M^\textrm{bary}$ indicates the overall average opinion, which
quantifies the overall drift. }  \label{fig:dbmhkbetweenchairs}
\end{figure}

Figures \ref{fig:dbmdwdrift} and \ref{fig:dbmdwdriftextreme} show density-based
dynamics for the DW model with an essentially uniform but perturbed initial
distribution which correspond to the examples in Figure \ref{fig:hetdwdrift}.
The same effects are visual. Finally, in
Figures \ref{fig:dbmhkdrift} and \ref{fig:dbmhkbetweenchairs} similar examples 
as in Figure \ref{fig:hethkdrift} are reproduced in the density-based HK model 

\

\section{Conclusions}
\label{sec:conclusions}
It was demonstrated that societies with open- and a
closed-minded agents can find consensus
even when both bounds of confidence are surprisingly low. This adds a new
phenomenon where diversity of agents has drastic effects.

The systematic simulations in Section \ref{sec:results} also shows that
the example runs shown in Section \ref{sec:models-with-heter} are
generic. Effects of heterogeneity are more drastic as
stated by Weisbuch et al \cite{Weisbuch.Deffuant.ea2002Meetdiscussand} for the
DW model where it is only
claimed that the dynamics of the higher $\eps$ will govern the evolution
of clusters in the long run. Here we see that effects are much larger, allowing
convergence to consensus when both bounds are low but different due to a subtle interplay.

The effect is to a large extent due to the symmetric initial
situation around the mean opinion. The symmetry is conserved during
dynamics therefore no overall drifts can occur. The examples in Section \ref{sec:drifting} show that severe drifts of the whole
opinion profile may occur even if the initial
distribution is random and essentially uniformly distributed. This gives rise to
the speculation that severe
drifting phenomena are also ubiquitous under heterogeneous bounds of
confidence. So, they need not only happen in stylized situation as in
\cite{Deffuant.Neau.ea2002HowCanExtremism,Amblard.Deffuant2004roleofnetwork,
Deffuant2006ComparingExtremismPropagation}.  Drifting of the mean
opinion also happens in opinion dynamics in the real
world. The interplay of clustering and drifting (pulled by open-minded
agents towards closed-minded agents) is also quite realistic in the
political realm. So, these theoretical results might help to uncover hidden dynamics in real world opinion dynamics, which in turn can help to design better communication systems. 

\bibliographystyle{unsrt}
\bibliography{refs}
\end{document}